\documentclass[prl,twocolumn,floatfix,showpacs,showkeys]{revtex4}
\usepackage{latexsym,amssymb,amsmath,amsfonts}
\usepackage{graphicx}
\usepackage{bm}
\usepackage{setspace}
\newcommand{\beq}{\begin{equation}}
\newcommand{\eeq}{\end{equation}}
\newcommand{\bc}{\begin{center}}
\newcommand{\ec}{\end{center}}
\newcommand{\eeqa}{\end{eqnarray}}
\newcommand{\beqa}{\begin{eqnarray}}
\newcommand{\no}{\noindent}

\newcommand{\na}{\nabla}

\newcommand{\ga}{\gamma}

\newcommand{\de}{\delta}

\newcommand{\la}{\lambda}

\newcommand{\si}{\sigma}

\newcommand{\ph}{\phi}

\newcommand{\ed}{\end{document} }
\begin{document}

\title{New spin on Einstein's non-symmetric metric tensor}
\author{Richard T. Hammond}

\email{rhammond@email.unc.edu }
\affiliation{Department of Physics\\
University of North Carolina at Chapel Hill\\
Chapel Hill, North Carolina and\\
Army Research Office\\
Research Triangle Park, North Carolina}

\date{\today}

\pacs{41.60.-m, 03.50.De}
\keywords{non-symmetric metric}

\begin{abstract}
A solution to the gravitational field equations based on a non-symmetric metric tensor is examined. Unlike Einstein's interpretation of electromagnetism, or Moffat's generalized gravity, it is shown that the non-symmetric part of the metric tensor is the potential of the spin field. This is in agreement with string theory and provides a natural coupling between gravitation and strings.
\end{abstract}

\maketitle

Einstein spent over three decades searching for, what he called, a Unified Field Theory.\cite{einstein} In his mind, Einstein envisioned a theory of gravitation and electromagnetism emanating from a single entity, the non-symmetric metric tensor (NMT), just as electricity and magnetism are unified and described by the single entity, the electromagnetic field tensor. Although his efforts bore little fruit, some of that work was revived in the 1970s --not as a theory of gravity and electromagnetism, but simply a generalized theory gravity.\cite{moffat} These theories had problems when viewed purturbatively and faded.\cite{damour} However, string theory called the NMT to the forefront because there is a natural coupling to the NMT. In an attempt to unearth the significance of this term, a simple solution is described below. It is shown the intrinsic spin of elementary particles is the origin of the antisymmetric part of the NMT.

The quiddity of gravitation is represented by the metric tensor and the affine connection, each of which has a geometrical meaning. The metric tensor gives the  interval between two nearby events, and the affine connection has an equally strong visual: It gives the change in a vector upon parallel transport. Such changes can arise due to curved space, or in flat space due to curvilinear coordinates. To filter out the effects of curvilinear coordinates, we consider the change in a vector as it is parallel transported around a small, closed path. In this case the change in the vector is proportional to the curvature tensor, which is the fundamental indicator of spacetime curvature. From the curvature tensor one constructs the curvature scalar, $R$, which becomes the Lagrangian of the action.\cite{einstein} 

In order to develop a geometrical theory, a relationship between these two fundamental quantities, the metric tensor and the affine connection, must be found. Once this is done the entire theory is nearly established! But one could ask, why tamper with Einstein's perduring 1915 theory of general relativity (which I will refer to as GR)? There are many observations that attest to its correctness, and in the coming years we expect to be using gravitational wave measurements to open a new window to the universe.\cite{ligo}

Yet there are several reasons to seek a more general theory. One that strikes us right away is its isolation from the standard model: It is an outcast, a hulking theory incapable of being tamed by quantum field theory. One hope is that in a more general formulation it would succumb to quantization. As an example, only after the weak force was combined with electromagnetism was the theory renormalizable. In the 1970s, and after, much effort went into forming local gauge theories of gravitation in hopes that, like other local gauge theories, a suitable quantum program could be developed. 

Equally important is this: Einstein used Riemannian geometry, which at that time was the most general geometry available, but soon after Einstein's 1915 publication, mathematicians had developed the full theory of curved space. In this context, we see that Riemannian geometry is only a small part of curved space, and if gravity is the physics of Riemannian geometry, what is the physics, one is led to ask, of the rest of the curved space? 

In addition is the advent of string theory, which calls for a generalization of GR for several reasons, as will be described below. Thus, it is important to look beyond GR, and seek what dwells in the shadows of unplumbed curvature.

 In Einstein's original theory of general relativity (GR) the metric tensor was assumed to be symmetric, i. e., $g_{\mu\nu}=g_{\nu\mu}$. There is no obvious reason to make this assumption, but it simplifies the mathematics, and the physics, considerably.
In GR there are two ways of finding the relationship between the metric tensor and the affine connection. One is to assume the covariant derivative of the metric tensor vanishes, which is called the metricity condition. This insures the length of a vector upon parallel transport is invariant.
As a counterexample, Weyl developed a theory in which the length of a vector--the potential of electromagnetism--was not invariant, but Einstein showed the theory was unphysical.\cite{weyl}

The other way to establish this relation is called the Palatini method. In this, besides varying the action with respect to the metric tensor (assuming the Lagrangian is not an explicit function of its derivatives), another set of equations is derived by also taking variations with respect to the affine connection. This establishes the mathematical relation between it and the metric tensor. In GR, these two approaches give identical results, but with a non-symmetric tensor they may not.


We start by following Einstein with the variation of the action $\de I$ where we take $I$ to be, as usual,

\beq\label{action}
I=\int d^4x\sqrt{-g}R
\eeq

\no where $R$ is the curvature scalar. Now we explore the consequences of Einstein's use of the metricity condition,

\beq\label{+-}
\na_\si g_{\mu \nu}=g_{\mu\nu},_\si-\Gamma_{\mu\si}^{\ \ \theta}g_{\theta\nu}
-\Gamma_{\si\nu}^{\ \ \theta}g_{\mu\theta}
\eeq

\no where $\Gamma_{\mu\si}^{\ \ \theta}$ is the affine connection and the antisymmetric part is called the torsion,\cite{hehl},\cite{tg}
\beq
S_{\mu\nu}^{\ \ \si}=\frac12\left(
\Gamma_{\mu\nu}^{\ \ \si}-\Gamma_{\nu\mu}^{\ \ \si}\right)
.\eeq

Also, since the metric tensor is not symmetric it can be written uniquely in terms of its
symmetric part $g_{(\mu\nu)}\equiv \ga_{\mu\nu}$  and the antisymmetric part $g_{[\mu\nu]}\equiv \ph_{\mu\nu}$ so 

\beq
g_{\mu\nu}=\ga_{\mu\nu}+\ph_{\mu\nu}
.\eeq

In GR, both $\ph_{\mu\nu}$ and torsion, $ S_{\mu\si}^{\ \ \theta}$, are assumed to be zero. Although a non-symmetric metric tensor has been assailed,\cite{damour} the torsion tensor has been predicted to give small but physical effects. For this reason we have a good physical interpretation of torsion: It arises from the intrinsic spin of elementary particles. Just as an electron produces a gravitational field due to its mass, and an electromagnetic field due to its charge, it produces a torsion field due to its spin. This will be described in more detail below. We shall not worry about the perturbative problem raised in \cite{damour}, which we know can be corrected.

The ability to form two distinct covariant derivatives is one of the difficulties introduced by a non-symmetric connection and metric. This means, besides, (\ref{+-}), there are three other combinations one could form. In addition there are two different curvature tensors. If one were to take a linear combination of all possibilities, the theory would be caught torrent of unknown constants, rendering the theory unnavigable. To overcome this difficulty, Einstein concocted the Hermitian Principle.\cite{hermitian} With this nostrum he was able to boil down the pot of choices to (\ref{action}) and (\ref{+-}).

With a metric we may raise and lower indices. However, with a non-symmetric metric this is not a unique process. In addition, since $\ga_{\mu\nu}$ is a tensor and has an inverse (we insist this exists) we can use this to define another kind of operation, which we denote with an underbar, $S_{\mu\nu{\underline\si}}\equiv
\ga_{\la\si}S_{\mu\nu}^{\ \ \la}$

In the following I will use Einstein's equations, (\ref{action}) and (\ref{+-}), but I will explore the significance of a particular solution for the antisymmetric part of the metric tensor. In fact, I will show this case gives rise to a new but pleasing physical interpretation of $\ph_{\mu\nu}$. In all of Einstein's Unified Field Theory work he assumed $\ph_{\mu\nu}$ was related to the electromagnetic field. Much later Moffat\cite{moffat} eschewed this assumption and took the new field to be just another part of gravity. Here I will show  $\ph_{\mu\nu}$ is the potential of the spin field, and its origin is the intrinsic spin of elementary particles. A complete theory of torsion that is derived from an antisymmetric potential has been reviewed in the literature.\cite{tg} This work will help us make the identification of the antisymmetric part of the metric tensor as being associated with spin.

 The main point of this paper is to show the antisymmetric part of the metric tensor is related to the intrinsic spin of elementary particles. To accomplish this, we shall be content to choose a reasonable Lagrangian density, such as (\ref{action}), and not worry about the fact there are other choices for the curvature tensor, as described above.
There is no way out of this difficulty. Einstein's Hermitian principle gives us a mathematical way to single out a particular combination of terms, but this is not a physical principle. After all, he introduced the complex quantities, which are unphysical, in order to accomplish this. More importantly, it has been shown different choices of the scalar lead to essentially the same results.\cite{tonnelat}

Thus, we  adopt (\ref{+-}) and find
 
 \beq\label{gamma}
 \Gamma_{\mu\nu}^{\ \ \si}=S_{\mu\nu}^{\ \ \si}
 +C_{\mu\nu}^{\ \ \si} +\ga^{\si\theta}
 (S_{\theta\mu}^{\ \ \la}\ph_{\la\nu}+S_{\theta\nu}^{\ \ \la}\ph_{\la\mu})
\eeq

\no where $C_{\mu\nu}^{\ \ \si} $ is the form of the Christoffel symbol but made with $\ga_{\mu\nu}$ and its inverse $\ga^{\mu\nu}$.

In addition, (\ref{+-}) places another restriction on $\ph_{\mu\nu}$. In this note we explore the consequences of looking at a particular solution of this other restriction, which is

\beq\label{fundsol}
S_{\mu\nu{\underline\si}}=\frac13\left(\ph_{\nu\mu,\si} +\ph_{\si\nu,\mu}+\ph_{\mu\si,\nu}\right)
.\eeq

In general, once we addopt  (\ref{+-}) we cannot further impose the Palatini method because the metric tensor and the affine connection are no longer independent.    
However, with (\ref{fundsol}) we find the Palatini variation automatically vanishes.

In the weak field limit the underbar in the above equation disappears and we are immediately drawn to physics. Part of what pushed Einstein and others who tried to unifiy electromagnetism and gravity must be the recognition that, in this limit, the vanishing of torsion leads to the two homogeneous Maxwell equations if $\ph_{\mu\nu}$ is the electromagnetic field tensor, an intriguing harbinger since the full theory of electromagnetism was the concupiscible desideratum, back then.

This path, however, led Einstein deep into a disappointing quagmire of mathematics. Moreover, the electromagnetic field is correctly associated with the weak force in the local gauge theory of the standard model, and certainly seems to be independent of gravity. {\it Today we recognize other physics in} (\ref{fundsol}). It has been used as the definition of torsion that arises from a potential.\cite{dt} In fact, a complete theory has been developed in which torsion arises from the intrinsic spin of elementary particles.\cite{stf} The equation of motion was found from the field equations, and it was shown that conservation of total angular momentum plus spin is conserved in the absence of forces.  In addition, this is seen to be the antisymmetric field, or Kalb Ramond,\cite{kalb} field of string theory, which was shown to be the torsion\cite{scherk} of spacetime.\cite{op}
None of these observations was available to Einstein and his collaborators, but now they stand out like a beacon. This is the key observation that rekindles interest in the non-symmetric field, but it is not the only one.

 The other observation unavailable during Einstein's time concerns the string coupling to matter. In the field equations of GR, the left hand side represents the curvature of space, while the right side is the distribution of mass-energy that curves the space. In order to have a testable theory, the right hand side must be articulated.

This is because the field equations predict the equation of motion, and the material Lagrangian must be developed so that its concomitant energy momentum tensor correctly describes the source while the conservation laws produce equation of motion. Now enters string theory, and it is most natural to use the material string Lagrangian to couple to the source. But there turns out to be two choices. One is called the Nambu-Goto coupling which couples to the symmetric part of the metric tensor, and the other is known as the Kalb-Ramond coupling, which is antisymmetric and can therefore only couple to an antisymmetric field.

However, when we examine gravity,  we find there are no natural second rank tensors in GR. However, when we go beyond GR we see we do have such an antisymmetric tensor. One is the torsion potential, and this has been discussed in detail elsewhere,\cite{sig} and the other natural tensor is the antisymmetric part of the metric tensor.

Now we find, with (\ref{fundsol}), these two are naturally creolized. We have a new and natural explanation of the metric tensor. In GR the (symmetric) metric tensor is considered to be the ``potential'' of the gravitational field, and the affine connection is the ``field.'' With the non-symmetric metric, we have that definition in tact, but may also say the antisymmetric part of the metric tensor is the potential of the spin field, and the new parts of the affine connection are the torsion field.

There is yet other evidence of the probity of this approach reaching back to the 1970s. It was shown that general relativity with torsion could be formulated as a local gauge theory under the Poincare group. Now, we know there are two Casimir invariants of this group, $P^2$ and $L^2$--the square of the translation operator and 
Pauli-Lubanski spin operator. Although this was formulated with a symmetric metric tensor, so was the spin theory resulting from the torsion potential.\cite{tg} In that case, the ideas not only carry over to the non-symmetric case, they provide an interpretation as well
as a raison d'tre for the non-symmetric metric tensor.  In any case, we have come to think of mass and spin as being associated with the gravitational field, and the above results show a very natural framework for just that.


With (\ref{gamma}) and (\ref{fundsol}) in hand, we may proceed to derive the field equation from the variational principle (\ref{action}).  In vacuum we may show the variations lead to

\beq
R_{\mu\nu}=0
.\eeq

\no From this we have both the symmetric part equals zero, $R_{(\mu\nu)}=0$ and the antisymmetric part equals zero, $R_{[\mu\nu]}=0$. The antisymmetric equation may be shown to boil down to, in the limit of weak fields,

\beq\label{tfe}
S_{\mu\nu\ ,\si}^{\ \ \si}=0
\eeq

\no {\it which is precisely the same} as that obtained in \cite{stf}. Solutions were found, and it was shown that (\ref{tfe}) admits a dipole solution.\cite{stf} The equation of motion was also derived and produced the correct conservation laws only if the source of (\ref{tfe}) arose from the intrinsic spin. This corroborates the interpretation that, for the work presented in this Letter, the antisymmetric part of the metric tensor arises from the intrinsic spin of elementary particles.

In order to introduce the source to this set of equations, it was already shown that the antisymmetric Kalb-Ramond string coupling does the trick.\cite{shs} It may be shown that in the weak field limit the same results carry through as for the previous theory. It is this fact, coupled with the results above, that allows us to interpret the antisymmetric part of the metric as being the potential of a spin field, the torsion.

The approach taken here may be summarized as follows. First, following Einstein, we use a non-symmetric metric tensor and assume (\ref{+-}) holds.  We examine a particular solution resulting from (\ref{+-}), i.e., (\ref{fundsol}), which leads to the equations given above. In this special case it turns out, like GR, the Palatini variation yields the same result as the variation with respect to the metric tensor.

In this approach there is a natural string coupling: The symmetric part of the metric tensor is associated with the Nambo-Goto term and the antisymmetric part with the Kalb-Ramond term. This gives a natural union between gravity and string theory.

The physical interpretation is that particles with intrinsic spin give rise to a new field. 
 For example, in hydrogen, the proton has spin and so does the electron. According to the physical interpretation, there is an interaction--not what we call the spin-spin interaction, which is really electromagnetic in origin--a true spin-spin interaction independent of electromagnetism. It has not been observed in hydrogen, but this may be used to place an upper limit on the coupling constant.\cite{ub} In fact, other experiments have been performed over the years establishing the best  value of the upper bound of the coupling constant.\cite{see}

In previous work \cite{tg} the spin field was represented solely by torsion, but {\it it had to be assumed} to have a potential. The present work answers two questions nicely, cementing the theory. One is, why is torsion derived from a potential? In fact, in local gauge theories I mentioned above, this is not the case, and the torsion is taken to be the potential. Now, from (\ref{fundsol}) we have the reason. This follows directly from the fundamental assumption (\ref{+-}). It also answers the other equation, why is the metric tensor non-symmetric? Because it is the potential of torsion.

This may sound circular, but it is not. The necessity of torsion, from a potential, has already been demonstrated in two ways. First it was in shown in \cite{stf} the correct law for the conservation of total angular momentum plus spin can only be achieved with torsion. This in itself is a strong enough argument for its presence, but it was also shown that it is necessary from gauge invariance arguments.\cite{-m}

In the final analysis then, we have particles with intrinsic spin giving rise to a field. The potential of this field is the antisymmetric part of the metric tensor. This is a new physical interpretation of the antisymmetric part of the metric tensor entirely  different from previous work.

Finally, it is pointed out that torsion has always been considered metempirical, to weak to be of interest. But in this formulation there are new terms, nonlinear in $\ph_{\mu\nu}$, that may produce large effects. Much more research needs to be done.

\ed